\documentclass[aps,showpacs,twocolumn,prl]{revtex4}

\IfFileExists{srcltx.sty}{\usepackage[active]{srcltx}}%

\usepackage{epsfig}
\usepackage{graphicx}
\usepackage{epsfig}
\usepackage{bm}
\usepackage{amsmath}
\usepackage{amssymb}
 
\newcommand{\beq}{\begin{equation}}
\newcommand{\eeq}{\end{equation}}
\newcommand{\bea}{\begin{eqnarray}}
\newcommand{\eea}{\end{eqnarray}}

\newcommand{\mh}{m_{_{\rm H}}}
\newcommand{\ms}{m_{_{\rm S}}}
\newcommand{\lh}{\lambda_{_{\rm H}}}
\newcommand{\ls}{\lambda_{_{\rm S}}}
\newcommand{\lm}{\lambda_{_{\rm HS}}}
\newcommand{\sh}[1]{#1\hskip-7pt \diagup}

\def\simle{\lower 2pt \hbox {$\buildrel < \over {\scriptstyle \sim }$}}
\def\simge{\lower 2pt \hbox {$\buildrel > \over {\scriptstyle \sim }$}}

\begin{document}

\preprint{UCLA/06/TEP/23}

\title{Sterile neutrinos, dark matter, and the pulsar velocities in models with
a Higgs singlet}

\author{Alexander Kusenko
}

\affiliation{Department of Physics and Astronomy, University of California, Los
Angeles, CA 90095-1547, USA }

%\date{}

\begin{abstract}

We identify the range of parameters for which the sterile neutrinos can
simultaneously explain the cosmological dark matter and the observed velocities
of pulsars. To satisfy all cosmological bounds, the relic sterile neutrinos
must be produced sufficiently cold.  This is possible in a class of models with
a gauge-singlet Higgs boson coupled to the neutrinos. Sterile dark
matter can be detected by the x-ray telescopes. The presence of the singlet 
in the Higgs sector can be tested at the Large Hadron Collider. 

\end{abstract}

\pacs{14.60.St, 95.35.+d,   \hfill UCLA/06/TEP/23}

\maketitle

Sterile neutrinos with masses of the order of a few keV can
explain the observed velocities of
pulsars~\cite{ks97,fkmp,Fryer,Kusenko:review}, can
play a role in the star formation and reionization of the
universe~\cite{reion}, and can be the cosmological dark
matter~\cite{dw,Fuller,dolgov_hansen,shi_fuller} if
their relic population is sufficiently ``cold''.  
% If the sterile neutrinos 
% do have masses at the keV scale, one should try to understand the origin
% of this new energy scale.  The underlying physics can provide new
% mechanisms for production and cooling of the relic neutrinos. 

While it is possible that the keV mass scale is a new
fundamental constant of nature, it is of
interest to consider the possibility that the sterile neutrino masses arise
from the Higgs mechanism, just like the masses of the other fermions. Then the
keV scale is generated by the Higgs vacuum expectation value (VEV), and it
depends on the coupling.  The
sterile neutrinos can couple to the SU$_3\times$SU$_2\times$U$_1$ singlet Higgs
boson $S$, whose vacuum expectation value can give them the Majorana masses. 
Models of this kind have been
proposed~\cite{roberto,shaposhnikov_tkachev}, and their implications for the
electroweak phase transition, baryogenesis, and collider searches have been
studied in detail~\cite{Enqvist:majoron}. The production of relic sterile
neutrinos has also been studied in one specific limit of parameters, in which
the singlet $S$ with a sub-GeV mass can play the role of the
inflaton~\cite{shaposhnikov_tkachev}.   Here we will
consider a more general range of parameters, focusing on the region in which
the sterile neutrinos can simultaneously explain pulsar kicks and dark
matter, while satisfying the Lyman-$\alpha$, x-ray, and other bounds.  We will
see that, if the scalar Higgs mass is of the order of the electroweak scale,
all these constraints can be satisfied simultaneously, and, in particular, the
momentum distribution of the relic sterile neutrinos can be cold enough for
dark matter.  The mixing between the Higgs doublet and the singlet can be
probed by the upcoming experiments at the LHC.

The neutrino masses can be introduced by means of the
following addition to the Standard Model lagrangian: 
\bea
{\cal L}
  & = &  {\cal L}_{0}+\bar N_{a} \left(i \sh{\partial} \right )
N_{a}
  - y_{\alpha a} H \,  \bar L_\alpha N_{a} \nonumber \\ 
& & - \frac{h_a}{2} \, S \, \bar {N}_{a}^c N_{a} 
 +V(H,S) + h.c. \,, 
\label{lagrangianS}
\eea
where $ {\cal L}_{0}$ includes the gauge and kinetic terms of the Standard
Model, $H$ is the Higgs doublet, $S$ is the real singlet, $L_\alpha$
($\alpha=e,\mu,\tau$) are the lepton doublets, and $ N_{a}$ ($a=1,...,n$) are
the additional singlet neutrinos.  Let us consider the following scalar
potential: 
\bea
V(H,S) & = &  \mh^2 |H|^2 + \ms^2 S^2+ 
\lm |H|^2 S^2+ \ls  S^4  \nonumber \\ &
 & + \lh |H|^4 .
\label{potential}
\eea

After the electroweak symmetry breaking, the Higgs doublet and singlet fields
each develop a VEV, $\langle H\rangle= v_0=247$~GeV, $\langle S\rangle= v_1$,
and the singlet neutrinos acquire Majorana masses $ M_a = h_a v_1$.   
% The
% Nambu--Goldstone boson associated with the lepton-number symmetry breaking is
% so
% weakly coupled that it escapes direct
% detection~\cite{roberto,Enqvist:majoron}. 
The masses of the Higgs doublet and singlet at zero temperature are  
\beq
\tilde{m}^2_{_{H,S}} = \lh v_0^2+\ls v_1^2 \pm \sqrt{D}, 
\label{masses}
\eeq
where $D= (\lh v_0^2 - \ls v_1^2)^2+\lm^2 v_0^2 v_1^2$.  Below the scale of
this symmetry breaking, the low-energy effective lagrangian is
\beq
{\cal L}
  = {\cal L_{\rm SM}}+\bar N_{a} \left(i \sh{\partial} \right )
N_{a}
  - y_{\alpha a} H \,  \bar L_\alpha N_{a} 
  - \frac{M_{a}}{2} \; \bar {N}_{a}^c N_{a} + h.c. \,,
\label{lagrangianM}
\eeq
where $ {\cal L_{\rm SM}}$ is the Standard Model lagrangian. This is the usual
seesaw lagrangian~\cite{seesaw}.  The number $n$  of singlet neutrinos is not
limited by the anomaly constraints or any other theoretical considerations,
and the experimental limits exist only for larger mixing
angles~\cite{sterile_constraints}.  Supernova 1987A provides a
constraint~\cite{Dolgov:2000pj,Fuller}, which depends on the mixing angle
and the sterile neutrino mass~\cite{Fuller}.  To explain the neutrino masses
inferred from the atmospheric and solar neutrino experiments, $n=2$ singlets are
sufficient~\cite{2right-handed}, but a greater number is called for if the
lagrangian (\ref{lagrangianM}) is to explain the
LSND~\cite{deGouvea:2005er}, the r-process nucleosynthesis~\cite{r}, the pulsar
kicks~\cite{ks97,fkmp,Fryer}, and dark
matter~\cite{dw,Fuller,dolgov_hansen,shi_fuller,nuMSM}. The scale of the
right-handed
Majorana masses $M_{a}$ is
unknown; it can be much greater than the electroweak scale~\cite{seesaw},
or it may be as low as 1-10~eV~\cite{deGouvea:2005er,deGouvea:2006gz}. The
seesaw mechanism~\cite{seesaw} can explain the smallness of the neutrino masses
in the presence of the Yukawa couplings of order one if the
Majorana masses are much larger than the electroweak scale; then 
the light neutrino masses are suppressed by the ratio $ \langle H \rangle/M$. 
However, the origin of the Yukawa couplings remains unknown, and there is 
no evidence that these couplings must be of order 1.  In fact, the Yukawa
couplings of the charged leptons are much smaller than 1. 
Theoretical {\em naturalness} arguments in favor of the 
low-energy seesaw~\cite{deGouvea:2005er} appear to be as compelling as those in
favor of the high-scale seesaw~\cite{seesaw}.  In both limits one can have a
successful leptogenesis: in the case of the high-scale seesaw, the baryon
asymmetry can be generated from the out-of-equilibrium decays of heavy
neutrinos~\cite{Fukugita:1986hr}, while in the case of the low-energy seesaw, 
the neutrino oscillations produce the asymmetry~\cite{baryogenesis}.

Let us first consider just one singlet neutrino, with
Majorana mass of the order of several keV, in light of its possible role in
explaining pulsar kicks and dark matter. The same neutrino can
play an important role in star formation and reionization of the
universe~\cite{reion}, as well as other astrophysical
phenomena~\cite{biermann_munyaneza}. 

\begin{figure}[ht]
%\epsfxsize=10cm   %width of figure - will enlarge/reduce the figures
% \epsfbox{sterile_limits.eps}
%\figurebox{2cm}{3cm}{} %to have a box alone 
\centerline{\epsfxsize=3.2 in\epsfbox{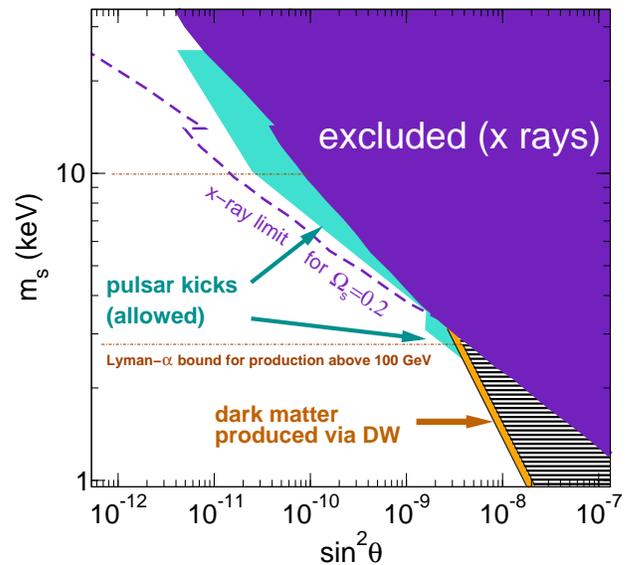}}   
\caption{The x-ray limits reported in Ref.\cite{x-rays} (dashed line) apply if
the sterile neutrinos account for all the dark matter ($\Omega_s=0.2$). 
The value of $\Omega_s$ depends on the production mechanisms, but it cannot be 
lower than the amount produced via DW mechanism~\cite{dw} (except for the
low-reheat scenarios~\cite{low-reheat,deGouvea:2006gz}).  The
model-independent exclusion plot (purple region) is obtained by assuming this
minimal value.  A sterile neutrino with mass 3~keV and $ \sin^2 \theta
\approx 3\times 10^{-9}$, produced at some temperature above 100 GeV, can
explain both pulsar kicks and dark matter.}
\label{figure:range}
\end{figure}

The range of parameters consistent with the explanation of pulsar kicks is
shown in Fig.~\ref{figure:range}.  It contains two regions
corresponding to the resonance oscillations~\cite{ks97} and off-resonance
oscillations~\cite{fkmp}.  The boundaries of these regions are defined by the
requirements of (i) the supernova energetics, including the SN1987A constraints,
(ii) adiabaticity and weak damping for resonant oscillations, (iii) sufficient
anisotropy to explain the observed pulsar
speeds~\cite{ks97,fkmp,Kusenko:review}. 

If the coupling $h$ in eq.~(\ref{lagrangianS}) is small enough, the sterile
neutrinos are out of thermal equilibrium at any time after inflation.  This is
the case if the annihilations $NN \rightarrow NN$, $NN \rightarrow  {\rm
scalars}$, etc.  are not fast enough to keep the sterile neutrinos in
equilibrium for temperatures $T$ in the range $\tilde{m}_{_{S}}<T<T_{\rm
reheat}$.   For $\tilde{m}\sim 1$~TeV and the reheat temperature $T_{\rm
reheat}<10^{16}$~GeV, the sterile neutrinos are out of equilibrium for $h
\ll 10^{-6}$~\cite{Enqvist:majoron}.    

The sterile neutrinos can be produced out of equilibrium in two
different processes.  First, since $S$ is in thermal equilibrium at high
temperature, sterile neutrinos are produced through decays $S\rightarrow
NN$~\cite{shaposhnikov_tkachev}.  Most of these
neutrinos are produced at temperatures of the order of
$\tilde{m}_{_{S}}\sim  (0.1-1)$~TeV. 
Second, at much lower temperatures, $T\sim 0.1$~GeV, the sterile neutrinos 
are produced from oscillations of active neutrinos, as in the
Dodelson-Widrow (DW) scenario~\cite{dw}. If the lepton asymmetry is
relatively large, the sterile neutrinos are produced much more
efficiently~\cite{shi_fuller}.  The lepton asymmetry of the universe is not
know, but strong upper bounds do exist~\cite{Dolgov:2002ab}. 

These two production mechanisms operate sequentially.  $S$ decays
are governed by the coupling $h$, while the production via DW mechanism depends
on mixing angles.  In the limit $ y_{\alpha i} v_0
\ll \min_i (M_i)$, these mixing angles are given by the usual seesaw
relations: 
%
% \beq
% \theta_{\alpha i} = \frac{y_{\alpha i} \langle H \rangle}{M_i}
$\theta_{\alpha i} = y_{\alpha i} \langle H \rangle/M_i$. 
% \label{mixing_angles}
% \eeq
For simplicity we will assume that only one sterile neutrino has mass of
several keV and that only one of the mixing angles is non-zero.  This mixing
angle is a function of both couplings, $y$ and $h$,
as well as $\tan \beta=v_0/v_1$:
\beq
\theta = \frac{y \, \langle H \rangle}{h \, \langle S \rangle} = \frac{y}{h}
\tan
\beta. 
\eeq

There are two limits which, in combination, appear to stymie the
DW proposal~\cite{dw} for sterile dark matter.  We will see that the
models with singlet Higgs bosons offer a way to reconcile these
seemingly contradictory bounds. 

One important limit on the abundance of the relic sterile neutrinos comes from
the x-ray observations.  The relic neutrinos can decay into one of the active
neutrinos and a photon.  Since this is a two-body decay, the photon energy is
equal one-half of the particle mass.  The flux of expected x-ray photons
depends on the decay rate, which is proportional to $(\sin^2
\theta)$~\cite{pal_wolf}. 
This flux also depends on the relic abundance of sterile neutrinos, which, in
turn, depends on the value of $h$ (for the $S\rightarrow NN$ production
at temperature
$T\sim \tilde{m}_{_{S}}$) and $\theta$ (for the production via neutrino mixing
at temperature $T\sim 0.1$~GeV). If the latter is the dominant production
mechanism, then the sterile neutrinos can be the dark matter in the range of
masses~\cite{Abazajian:2005gj} below 3.5~keV, as shown in
Fig.~\ref{figure:range}.  For larger masses, the x-ray limits on the decay 
photons~\cite{x-rays} disallow the relic sterile neutrinos, unless their
abundance is smaller than what is required for dark matter.  However, if $S$
decays at $T\sim (0.1-1)$~TeV are the dominant source of sterile neutrinos,
they can still make up the entire dark matter, even if $\theta$ is arbitrarily
small. 

Another limit on dark matter comes from the Lyman-$\alpha$
observations~\cite{viel} and is based on the requirement that the dark matter
be cold enough to generate the smallest structures observed in the
absorption spectra of distant quasars.  This limits the free-streaming length
of dark matter, whose relation with the particle mass depends on the production
mechanism.  If the sterile neutrinos are produced via mixing, then the
Lyman-$\alpha$ bound is $m>10$~keV~\cite{viel}.  Obviously, this limit is in
conflict with the x-ray limit mentioned above, namely $m<3.5$~keV.  

If the sterile neutrinos account for only a part of dark matter, then the
Lyman-$\alpha$ bounds do not apply, and the x-ray bounds are weaker.
The x-ray limits~\cite{x-rays}, shown as a dashed line in
Fig.~\ref{figure:range},  are based on the assumption that the sterile
neutrinos constitute the entire dark matter, i.e.,
$\Omega_s = 0.2$. However, if $h\rightarrow 0$, this condition
is not satisfied for points along the dashed line; in fact, $\Omega_s \ll 0.2$
for most of these points.   This is because, in the absence of $S$ decays, the
density of sterile neutrinos in the universe may be too small to explain dark
matter.  The x-ray signal is correspondingly smaller in this case, and the
x-ray bounds are weaker.   However, there still exists an x-ray bound on a
sterile neutrino with a given mass and mixing angle, although this particle may
be only a small part of dark matter.  

The production via neutrino mixing \cite{dw}
provides the lowest possible abundance of sterile
neutrinos, as long as the universe was reheated to a GeV or higher
temperature. In low-reheat cosmological scenarios the bounds are
relaxed considerably~\cite{low-reheat,deGouvea:2006gz}.  Here we will assume
that the universe has reached temperatures well above TeV after inflation.
In Fig.~\ref{figure:range} we show both the bounds based on the
assumption $\Omega_s=0.2 $ and the model-independent exclusion region (solid
purple) based on the production only via the DW mechanism~\cite{dw}, in which
case $\Omega_s $ can be smaller than 0.2.  To calculate this production, we used
an analytical fit to the numerical calculation of
Abazajian~\cite{Abazajian:2005gj}. There may be some hadronic 
uncertainties in this calculation~\cite{Shaposhnikov_Laine}.

There is a range of parameters for which the sterile
neutrinos can explain the pulsar velocities and can affect the star formation,
although they may not be the dominant component of the dark matter
(see Fig.~\ref{figure:range}). However, it
is also possible to explain pulsar kicks and the dark matter
simultaneously.  Indeed,
if the $S$-decay mechanism dominates the production of relic neutrinos, they
may be redshifted sufficiently as the universe cools from $T_{_S}\sim
(0.1-1)$~TeV to $T < \, $MeV. 

If all the dark matter is made up of sterile neutrinos, the bound $m_s
>10$~keV~\cite{viel} applies to the DW scenario, in which the average
momentum of a keV neutrino at low temperature $T$ is
%\beq
$\langle p_s \rangle_{\rm DW} \approx 2.8 \, T$~\cite{Abazajian:2005gj}.
%\label{p_s_DW}
%\eeq
%
Any additional redshifting of sterile neutrinos (e.g., due to some 
entropy production~\cite{Asaka:2006ek}) relaxes the 10~keV limit to a lower
value. 

Production of sterile neutrinos via decay $S \rightarrow NN$ occurs mainly at
temperature $T_{_S}\approx \tilde{m}_{_{S}}$~\cite{shaposhnikov_tkachev}.  Our
model has two mass eigenstates given by eq.~(\ref{masses}), subject to thermal
corrections.  In the case of different  masses and non-negligible mixing, one
of the mass eigenstates decouples before the other. Each mass eigenstate has a
non-zero $S$ component and, therefore, contributes
to the production of the relic population of sterile neutrinos.  If we make a
simplifying assumption of equal masses (achieved when $\lh \langle H
\rangle ^2 \approx
\ls \langle S \rangle ^2 $ and when $\lm$ is small), the results of
Ref.~\cite{shaposhnikov_tkachev} are directly applicable to our case, and one
can write that the amount of sterile dark matter as
\bea
\Omega_s & = & 0.2 \left ( \frac{33}{\xi} \right )
\left ( \frac{h}{ 1.4 \times 10^{-8} } \right )^3
\left ( \frac{ \langle S \rangle }{\tilde{m}_{_S} } \right )=
\label{Omega_w_VEVs} \\ & & 
0.2 \left ( \frac{33}{\xi} \right )
\left ( \frac{h}{ 1.4 \times 10^{-8} } \right )^3 
\left ( \frac{1}{\sqrt{2 \lambda_{_S}}}\right ),
\label{Omega_w_lambda}
\eea
where $ \xi $ is the change in the number density of sterile neutrinos
relative to $T^3$ due to the dilution taking place as the universe cools down
from $T_{_{S}}\sim (0.1-1)$~TeV to a temperature well below
MeV. The sterile neutrinos produced in these decays have an almost thermal
spectrum at the time of production. More precisely,
their average momentum $\langle p_s \rangle = \pi^6/(378 \zeta(5)) T \approx
2.45 \, T$~\cite{shaposhnikov_tkachev}, while the average momentum of the 
relativistic fermions in equilibrium is $p_{_T}\approx 3.15\, T$.  As the
universe cools down, the number of effective degrees of freedom decreases from
$g_*(T_{_{S}})=110.5$ to $g_*(0.1\, {\rm MeV})=3.36$. Then $
\xi =g_*(T_{_S})/g_*(0.1\, {\rm MeV})\approx 33$. This
causes the redshifting of $\langle p_s \rangle$ by the factor $\xi^{1/3}$: 
\beq
\langle p_s \rangle_{(T \ll 1{\rm MeV})} = 0.76 \, T \left [ \frac{110.5}{
g_*(\tilde{m}_{_{S}})
}\right ]^{1/3}
\label{p_s_redshifted} 
\eeq
Comparing eq. (\ref{p_s_redshifted}) with the DW case, one concludes that,
as long as the population of sterile neutrinos is dominated by those produced
at high temperature (large enough $h$, small $\theta$), the Lyman-$\alpha$
limit changes from 10~keV to 
\beq
m_s> 2.7 \, {\rm keV}
\eeq 
This lower bound is shown in Fig.~\ref{figure:range}. 

In addition, the electroweak phase transition in this model can be
first-order, and the entropy production can further redshift
the momentum distribution of sterile neutrinos.  Enqvist et
al.~\cite{Enqvist:majoron} have found that the energy
density increase due to the phase transition can be at most $10\, T_0^4$,
where $T_0$ is the transition temperature.  This changes $\xi$ by at most a
factor 1.3.  

The presence of the singlet Higgs boson has important implications for the
Higgs searches.  Although we have focused so far on the keV sterile neutrino,
the present data requires $n \ge 3$ sterile neutrinos 
in eq.~(\ref{lagrangianS})~\cite{deGouvea:2005er,nuMSM}.  This is not in
conflict with big-bang nucleosynthesis (BBN): one or two additional thermalized
neutrinos are consistent with the BBN constraints at one (two) sigma
level~\cite{bbn}; besides some of these sterile neutrinos can be out of
equilibrium.  The couplings $h_a$
of the additional sterile neutrinos need not be small, and the mixing between
$H$ and $S$ can also be large.  Hence, the Higgs boson can decay invisibly. 
This possibility has the effect of
weakening the LEP bound on the mass of the lightest
Higgs~\cite{Enqvist:majoron,Binoth:1996au}. However, one can discover the 
invisible Higgs $h_{\rm inv}$ at the LHC in the $Z+h_{\rm inv}$
channel~\cite{Davoudiasl:2004aj}.  For Higgs mass of 120~GeV, the discovery is
possible at the LHC already with 10~fb$^{-1}$ in the $Z+h_{\rm inv}$ channel,
while 100~fb$^{-1}$ of data can afford the discovery in the weak boson fusion
channel~\cite{Davoudiasl:2004aj}.  For some range of couplings, the singlet
$S$ can decay into the visible channels and can be discovered  via displaced
vertices~\cite{Strassler}. 

To summarize, the relic sterile neutrinos with mass of a few keV can
simultaneously explain the pulsar velocities and dark matter if their
production in the early universe occured above the electroweak scale. 
Over a broader range of parameters 
(Fig.~\ref{figure:range}), the sterile neutrinos can explain pulsar kicks
and can play a role in the star formation, while they may not be the dominant
component of dark matter.

The author thanks M.~Shaposhnikov and I.~Tkachev for discussions. This work was
supported in part by the DOE grant DE-FG03-91ER40662 and by the NASA ATP grants
NAG~5-10842 and NAG~5-13399.  The author thanks CERN and EPFL for
hospitality.

\end{document}